\documentclass[aps,prl,twocolumn,english,superscriptaddress,longbibliography]{revtex4-1}
\usepackage{amsmath,amssymb,amsfonts}
\usepackage{CJK}
\usepackage{bm}
\usepackage{tikz}
\usepackage{babel}
\usepackage{color}
\usepackage{graphicx}

\usepackage{xspace}
\usepackage{epstopdf}
\usepackage{dcolumn}
\usepackage{longtable}
\usepackage{multirow}
\usepackage[colorlinks=true, letterpaper=true, pdfstartview=FitV, linkcolor=blue, citecolor=blue, urlcolor=blue]{hyperref}

\def \H {\mathcal{H}}
\def \K {\hat{\mathcal{K}}}

\def \I {\hat{I}}

\def \i {{\mathrm{i}}}

\def \Z {\mathbb{Z}}

\def \k {\bm{k}}

\begin{document}

\title{$\Z_2$-projective translational symmetry protected topological phases}

\author{Y. X. Zhao}
\email[]{zhaoyx@nju.edu.cn}
\affiliation{National Laboratory of Solid State Microstructures and Department of Physics, Nanjing University, Nanjing 210093, China}
\affiliation{Collaborative Innovation Center of Advanced Microstructures, Nanjing University, Nanjing 210093, China}

\author{Yue-Xin Huang}
\affiliation{Research Laboratory for Quantum Materials, Singapore University of Technology and Design, Singapore 487372, Singapore}

\author{Shengyuan A. Yang}
\affiliation{Research Laboratory for Quantum Materials, Singapore University of Technology and Design, Singapore 487372, Singapore}

\begin{abstract}
Symmetry is fundamental to topological phases. In the presence of a gauge field, spatial symmetries will be projectively represented, which may alter their algebraic structure and generate novel physics. We show that the $\Z_2$ projectively represented translational symmetry operators adopt a distinct anti-commutation relation. As a result, each energy band is twofold degenerate, and carries a varying spinor structure for translation operators in momentum space, which cannot be flattened globally.
Moreover, combined with other internal or external symmetries, they give rise to exotic band topologies. Particularly, with the inherent time-reversal symmetry, a single fourfold Dirac point must be enforced at the Brillouin zone corner.
By breaking one primitive translation, the Dirac semimetal is shifted into a special topological insulator phase, where the edge bands have a M\"{o}bius twist. Our work opens a new arena of research for exploring topological phases protected by projectively represented space groups.
\end{abstract}
\maketitle

{\color{blue}\textit{Introduction}.} Symmetry is of fundamental importance in physics. This is particularly manifested in the development of topological phases of matter. Initiated with the study of quantum Hall effects~\cite{Klitzing1980,Thouless1982,Haldane1988}, topological phases have expanded into a large family via the consideration of various symmetries~\cite{Chiu2016}: firstly the internal symmetries~\cite{Kane2005a,Schnyder2008,Kitaev2009,Zhao2013b}, such as time-reversal and particle-hole symmetries, and more recently the
crystal space group symmetries~\cite{Slager2013,Zhao2016a,Shiozaki2016,Bradlyn2017,Tang2019,Zhang2019,Vergniory2019,Tang2019a}. The symmetry group dictates the topological classification, restricts band topological features, and protects novel types of excitations.


Regarding symmetries, a very crucial yet often overlooked point is that: Physical systems in fact represent symmetry groups \emph{projectively}~\cite{moore_Abstract_Group}.
As the most elementary example, the time reversal symmetry $T$ generates the $\Z_2$ group with $T^2=1$. However, for particles with spin-$1/2$, $T$ is projectively represented to satisfy $T^2=-1$. Such distinct algebra arising from the projective representation is at the heart of the $T$-invariant topological phases, such as the quantum spin Hall insulators~\cite{Hasan2010,Qi2011}.

Now a natural question is: How about projectively represented space group (PRSG) symmetries? PRSGs are ubiquitous for both classical and quantum systems, as they generally appear in the presence of gauge degrees of freedom. However, their impact on the topological phases has not been studied before.

In this Letter, we investigate the most fundamental PRSG --- the projectively represented translation group (PRTG). Translational group is what defines a crystal, and is contained in all space groups. Here, we focus on its $\Z_2$ projective representation, motivated by noting that $\Z_2$ gauge fields emerge in a wide range of interesting systems. For example, the $\Z_2$ group is the remaining gauge group after Cooper pair condensation in superconductors~\cite{Discrete_Gauge,Superconductor_Z2,Xiao-Gang_RMP}. Many spin liquids have emergent $\Z_2$ gauge fields in the vicinity of their ground
states~\cite{Kitaev2006,Savary2016,Xiao-Gang_RMP,Zhao2020}. Moreover, it is supported by almost all $T$-invariant artificial periodic systems, such as photonic/phononic crystals~\cite{Lu2014,Yang2015,Photonic_Crystal_quadrupole,Acoustic_Crystal}, electric-circuit arrays~\cite{Imhof2018,Yu2020}, and mechanical networks~\cite{Huber2016,Prodan_Spring}, which are briefly discussed in the Supplemental Materials (SM)~\cite{Supp}.
The physics discussed in this Letter can be naturally realized in these systems.

We show that according to the second group cohomology, there is a unique nontrivial $\Z_2$ PRTG in two dimensions, for which the two translation generators anti-commute rather than commute with each other. A significant consequence is that each energy band must have a twofold degeneracy to projectively represent the translational symmetry. Thereby, for each isolated band the translational operators form a varying spinor structure, which cannot be flattened smoothly over the whole Brillouin zone (BZ) because of a nontrivial winding number. Moreover, interesting topological phases can be generated from this distinct algebra. We demonstrate that together with the inherent $T$-symmetry, the PRTG enforces a fourfold degenerate Dirac point at the corner of the BZ.
Furthermore, by breaking one primitive translation, e.g., via dimerization along one direction, the critical Dirac semimetal state can be transformed into a topological insulator phase protected by the other preserved primitive translation and the sublattice symmetry. The resulting topological insulator is characterized by a $\Z_2$ topological invariant, and features topological edge bands with a M\"{o}bius twist at any edge along the preserved translation.

{\color{blue}\textit{$\Z_2$ projective translational symmetry}.} We start with the basics of $\Z_2$ projective representations of the translation group.
Let $L_{\bm{a}_1}$ and $L_{\bm{a}_2}$ be the two generators of the translation group in two dimensions. They are defined by their action in real space:
$
L_{\bm{a}_{1,2}}\bm{r}=\bm{r}+\bm{a}_{1,2}
$, with $\bm{a}_{1,2}$ the two primitive lattice vectors.
Each of them generates a free Abelian group $\Z$, and they commute with each other
\begin{equation}
[L_{\bm{a}_1},L_{\bm{a}_2}]=0.
\end{equation}
Therefore, the translation group is isomorphic to $\Z\times\Z$.

As mentioned, the group will be projectively represented for physical systems, e.g., in lattice gauge theory. If we consider the $\Z_2$ gauge group, the $\Z_2$ PRTG then corresponds to the short exact sequence~\cite{moore_Abstract_Group},
\begin{equation}
0\rightarrow  \Z_2  \xrightarrow{i}  \mathcal{G} \xrightarrow{p} \Z\times\Z\rightarrow 0,
\end{equation}
where $\mathcal{G}$ is the extension of translation group by $\Z_2$, $i$ is the natural injection and $p$ is a projection. The possible extensions can be solved by using the fact that the short exact sequence is equivalent to the second group cohomological class~\cite{moore_Abstract_Group}
\begin{equation}
H^{2}(\Z\times\Z,\Z_2)\cong \Z_2.
\end{equation}
Thus, there are two classes of projective representations.
For the \emph{nontrivial} representation, in the extended group, the preimages of the generators, $\mathsf{L}_{\bm{a}_{1,2}}$, will satisfy the \emph{anti-commutation} relation
\begin{equation}
\{\mathsf{L}_{\bm{a}_{1}},\mathsf{L}_{\bm{a}_{2}}\}=0,
\end{equation}
or alternatively,
\begin{equation}\label{flux}
\mathsf{L}_{\bm{a}_{2}}^{-1}\mathsf{L}_{\bm{a}_{1}}^{-1}\mathsf{L}_{\bm{a}_{2}}\mathsf{L}_{\bm{a}_{1}}=-1.
\end{equation}
Note that the left-hand side of (\ref{flux}) moves a particle around the edges of a plaquette formed by the primitive vectors. The minus sign on the right hand side indicates that the particle acquires a $\pi$-phase in this process. Hence, this nontrivial projective representation can be achieved on a lattice where each plaquette has a $\pi$-flux (assuming the particle carries a unit charge). Actually, for particle-hole invariant interacting fermionic systems, such flux configuration is favored by the ground state~\cite{Lieb_Theorem}.

\begin{figure}
	\includegraphics[scale=0.55]{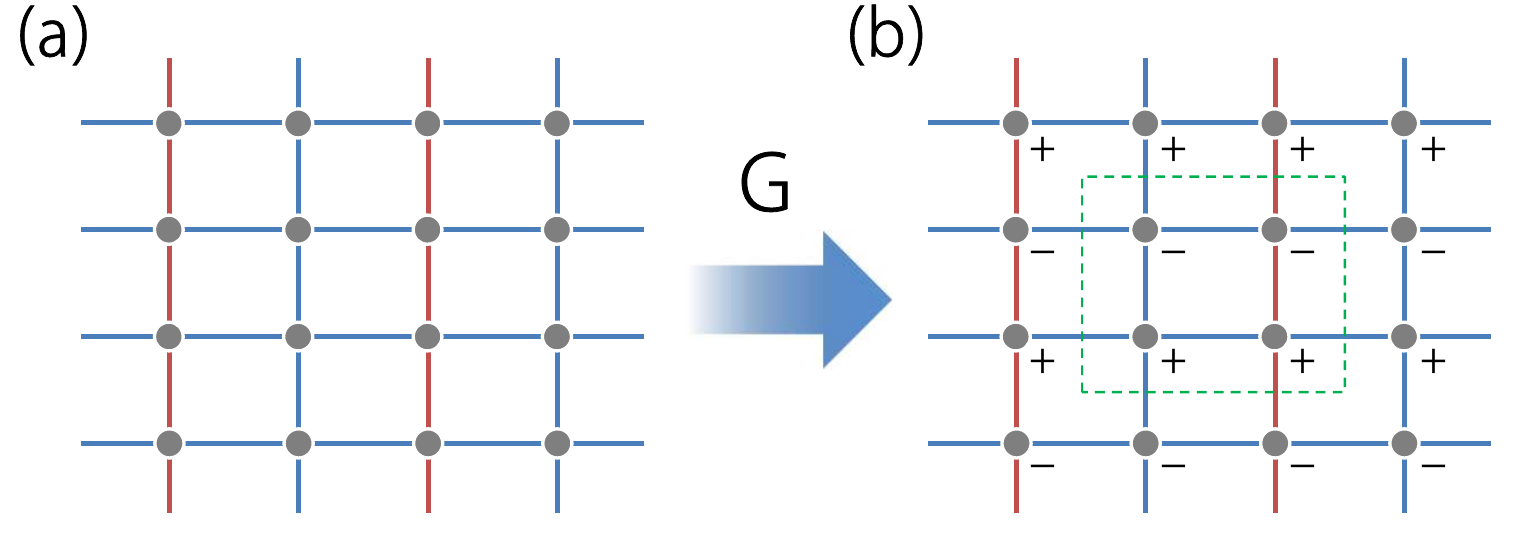}
	\caption{(a) Rectangular lattice with $\pi$-flux per plaquette.
 The gauge condition is chosen as that each red/blue bond has a negative/positive hopping amplitude. (b) Under the $\Z_2$ gauge transformation $\mathsf{G}$, sites in odd (even) rows are multiplied with a $\pi$ ($0$) phase. The dashed box indicates the selected unit cell.\label{fig:Rectangular_lattice}}
\end{figure}
{\color{blue}\textit{Proper translation in momentum space}.}
Without loss of generality, let us consider a rectangular lattice. As discussed, to have a nontrivial $\Z_2$ PRTG, we need each plaquette of the lattice has a $\pi$-flux. Figure~\ref{fig:Rectangular_lattice}(a) shows one simplest gauge configuration, where the hopping amplitudes along $y$ alternate in sign among the columns. After fixing this particular gauge configuration, one observes that the primitive translation $L_y$ is preserved, whereas $L_x$ is not manifestly preserved. To recover the original gauge pattern, we need to incorporate an additional gauge transformation $\mathsf{G}$, as illustrated in Fig.~\ref{fig:Rectangular_lattice}(b), which adds a $\pi$-phase for the sites in odd rows.
Namely, under the gauge condition, the \emph{proper} primitive translation operator along $x$ should be $\mathsf{L}_x=\mathsf{G}L_x$.  Meanwhile, we note that
\begin{equation}
\{\mathsf{G},L_y\}=0,
\end{equation}
because $L_y$ exchanges odd and even rows, while $\mathsf{G}$ alternates among the rows. Hence, 
\begin{equation}\label{Anti-commute}
\{\mathsf{L}_x, \mathsf{L}_y\}=0,
\end{equation}
which is consistent with our previous discussion (here, $\mathsf{L}_y=L_y$).

Let's proceed to consider the representations of the operators in momentum space. For this purpose, we need to first select an appropriate unit cell. Under the gauge configuration, the primitive cell consists of two sites, which are nearest neighbors in a row. However, the gauge transformation $\mathsf{G}$ does not respect the primitive cell. Hence, a proper unit cell should contain four sites, as illustrated in Fig.~\ref{fig:Rectangular_lattice}(b). Let $\tau_\mu$ and $\sigma_\mu$ be two sets of Pauli matrices operating on the row index and column index, respectively ($\mu=0,1,2,3$, and $\sigma_0=\tau_0=1_2$). Then, $L_x$ and $L_y$ are represented by
\begin{equation}
\hat{L}_x=\tau_0\otimes \left[\begin{matrix}
0 & 1\\
e^{\i k_x} & 0
\end{matrix}\right],\quad \hat{L}_y=\left[\begin{matrix}
0 & 1\\
e^{\i k_y} & 0
\end{matrix}\right]\otimes\sigma_0.
\end{equation}
The phase factor $e^{\i k_x}$ appears in $\hat{L}_x$, because the right column in one unit cell is translated under $L_x$ into the next cell. $e^{\i k_y}$ in $\hat{L}_y$ has the similar origin.
The gauge transformation $\mathsf{G}$ is represented by
\begin{equation}
\hat{\mathsf{G}}=\tau_3\otimes\sigma_0.
\end{equation}
Hence, the proper translation operators are given by
\begin{equation}\label{G-L-operators}
\hat{\mathsf{L}}_x=\hat{\mathsf{G}} \hat{L}_x= \tau_3\otimes \left[\begin{matrix}
0 & 1\\
e^{\i k_x} & 0
\end{matrix}\right],\quad \hat{\mathsf{L}}_y=\left[\begin{matrix}
0 & 1\\
e^{\i k_y} & 0
\end{matrix}\right]\otimes\sigma_0,
\end{equation}
consistent with the anti-commutation relation in \eqref{Anti-commute}.

{\color{blue}\textit{Symmetry-protected twofold degeneracy}.} An immediate consequence of PRTG is that each band must be twofold degenerate for every $\bm{k}$. Since $[\hat{\mathsf{L}}_x,\H(\bm{k})]=0$, let $\psi(\bm{k})$ be the simultaneous eigenstate of $\hat{\mathsf{L}}_x$ and $\H(\bm{k})$, namely, $\H(\bm{k})\psi(\k)=\mathcal{E}(\bm{k})\psi(\k)$ and $\hat{\mathsf{L}}_x\psi(\k)=\pm e^{\i k_x/2}\psi(\k)$. Because $\{\hat{\mathsf{L}}_x,\hat{\mathsf{L}}_y\}=0$, $\hat{\mathsf{L}}_y\psi(\k)$ is also an eigenstate of $\hat{\mathsf{L}}_x$ but with the opposite eigenvalue $\mp e^{\i k_x/2}$. Therefore, $\psi(\k)$ and $\hat{\mathsf{L}}_y\psi(\k)$ are orthogonal. Moreover, since $[\hat{\mathsf{L}}_y,\H(\bm{k})]=0$, $\hat{\mathsf{L}}_y\psi(\k)$ is also an eigenstate of $\H(\bm{k})$ with the same energy $\mathcal{E}(\k)$. Thus, under PRTG, each energy band must have a twofold degeneracy. This special feature is similar to $PT$-invariant systems with spin-orbit coupling (SOC), but clearly from a completely different physical origin here.

For an isolated twofold degenerate band, the two operators $\hat{\mathsf{L}}_{x,y}$ are, respectively, represented by
\begin{equation}\label{sigma-operators}
\Sigma_x(\k)=e^{\i k_x/2}\sigma_1,\quad
\Sigma_y(\k)=e^{\i k_y/2}\sigma_2,
\end{equation}
with $\{\Sigma_x(\k),\Sigma_y(\k)\}=0$. Actually, it is impossible to let both $\Sigma_x(\k)$ and $\Sigma_y(\k)$ be periodic in the BZ.  Therefore, $\Sigma_x(\k)$ and $\Sigma_x(\k)$ form a varying spinor structure in the BZ. The spinor structure cannot be flattened over the entire BZ due to the twist of the eigenstates over the large circle.  To see this, let us consider an adiabatic evolution of the eigenstate $\psi(\k)$ with $k_x$ increased by $2\pi$. The eigenvalue of $\hat{\mathsf{L}}_{x}$ experiences a counterclockwise rotation of $\pi$ on the complex plane, and therefore is connected to $\hat{\mathsf{L}}_y\psi(\k)$ in the end. This twist corresponds to the nontrivial winding number of the symmetry operator:
\begin{equation}
N=\frac{1}{2\pi i }\oint dk_x ~\mathrm{tr} \Sigma^\dagger_x(\k)\partial_{k_x}\Sigma_x(\k),
\end{equation}
which equals $1$. Parallel discussions can be made for $\Sigma_y(\k)$. Note that although operators \eqref{sigma-operators} are not periodic, a unitary transformation does not change the winding number.


\begin{figure}
	\includegraphics[width=8.5cm]{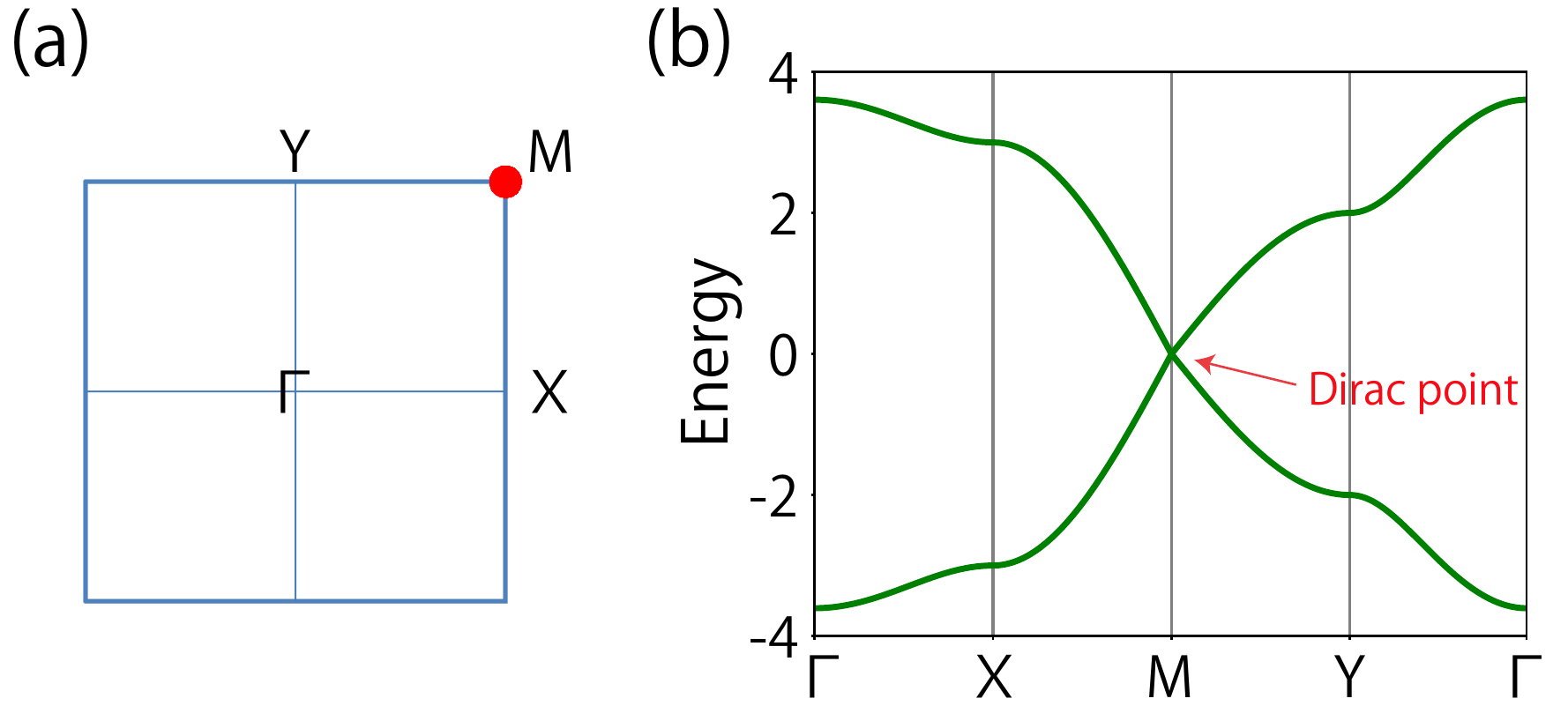}
	\caption{(a) Brillouin zone for the rectangular lattice. A fourfold Dirac point is enforced by symmetry at $M$ point (see main text). (b) shows the calculated band structure for model (\ref{HH}). Here, we take $t=1$, $J_1=J_2=1.5$.\label{fig:DP}}
\end{figure}

{\color{blue}\textit{Symmetry-enforced Dirac point}.} We first point out that the $\Z_2$ gauge theory preserves an inherent time-reversal ($T$) symmetry, which is represented by
\begin{equation}\label{T-operator}
\hat{T}=\K\I,
\end{equation}
for systems without SOC, where $\K$ denotes the complex conjugation and $\I$ the inversion of momenta. This is because, although $T$ reverses flux, the flux $\pi$ through a plaquette is equivalent to $-\pi$  for the $\Z_2$ gauge theory.



Interestingly, the PRTG, together and $T$, enforces a fourfold degenerate Dirac point at $M=(\pi,\pi)$ (momenta are in units of the respective reciprocal lattice constants).
To see this, one notes that at $M$, the two translation operators are given by
\begin{equation}
\hat{\mathsf{L}}_x^{M}= -\i \tau_3\otimes\sigma_2,\quad  \hat{\mathsf{L}}_y^{M}=-\i \tau_2\otimes\sigma_0.
\end{equation}
The Hamiltonian at the point $M$, $\H_M$, must be a \emph{real} matrix to preserve $T$, and commute with $\hat{\mathsf{L}}_{x,y}^{M}$. To analyze the form of $\H_M$, let us introduce $\gamma^1=\i \hat{\mathsf{L}}_x^{M}$ and $\gamma^2=\i \hat{\mathsf{L}}_y^{M}$, and extend them into a complete set of Hermitian Dirac matrices: $\gamma^3=\tau_3\otimes\sigma_1$, $\gamma^4=\tau_1\otimes\sigma_0$ and $\gamma^5=\tau_3\otimes\sigma_3$. Then, $\{\gamma^\mu,\gamma^\nu\}=2\delta^{\mu\nu}1_4$, with $\mu,\nu=1,2,\cdots,5$. The real linear space of $4\times 4$ Hermitian matrices is $16$-dimensional, and an orthogonal basis can be given by $1_4$, $\gamma^\mu$, and $\i\gamma^\mu\gamma^\nu$ with $\mu<\nu$. The subspace of Hermitian matrices commuting with both $\gamma^1$ and $\gamma^2$ has a basis: $1_4$, $\i\gamma^3\gamma^4$, $\i \gamma^3\gamma^5$ and $\i\gamma^4\gamma^5$. Hence, $\H_M$ is a linear combination of these four basis vectors. However, all of them are purely imaginary except $1_4$. Thus, $\H_M=\lambda 1_4$ and therefore is fourfold degenerate. Furthermore, the above argument actually asserts that the Hilbert space at $M$ should be decomposed into a direct sum of $4$D irreducible representations of symmetries of $\mathsf{L}_{x,y}$ and $T$, and accordingly the number of bands is a multiple of $4$.

Generically, for a $4$-band model, deviating from $M$, the spectrum  is gapped without symmetry protection. Hence, the Fermi point at $M$ is an isolated Dirac point (see Fig.~\ref{fig:DP}), solely guaranteed by PRTG and the inherent $T$ symmetry. The significance of the modified algebra for PRTG [Eq.~(\ref{Anti-commute})] is clearly demonstrated here, as such a Dirac point is not possible if the two translation operators commute. For a standard Dirac point, the bands around it are pairwise degenerate. For systems with SOC, this is usually achieved by the spacetime inversion symmetry $PT$ with $(PT)^2=-1$~\cite{Young2012}. Here, for the spinless fermions, the twofold degeneracy is resulted from the PRTG as aforementioned.


{\color{blue}\textit{M\"{o}bius topological insulator}.} The Dirac semimetal discussed above can be viewed as
a critical state. By selectively breaking the protecting symmetry, the state can transition into different topological phases.  Moreover, we observe that the translation operators acquire a particular momentum dependence, resembling that of twofold nonsymmorphic operators~\cite{Zak-Nonsymmorphic,Zhao2016}.
Since nonsymmorphic symmetries are well known for their induced band topology~\cite{Young2015,Shiozaki2015,Wang2016a,Watanabe2016,Bzdusek2016,Chang2017e,Wu2018c,Yu2019b,Zhang2020}, by revealing the common features, one can expect that PRTG will also generate rich topological phases in $\Z_2$ gauge systems.
Below, we show that breaking one primitive translation while maintaining the other will transform the system into a topological insulator with M\"{o}bius-twist edge bands.

\begin{figure}
	\includegraphics[width=8.6cm]{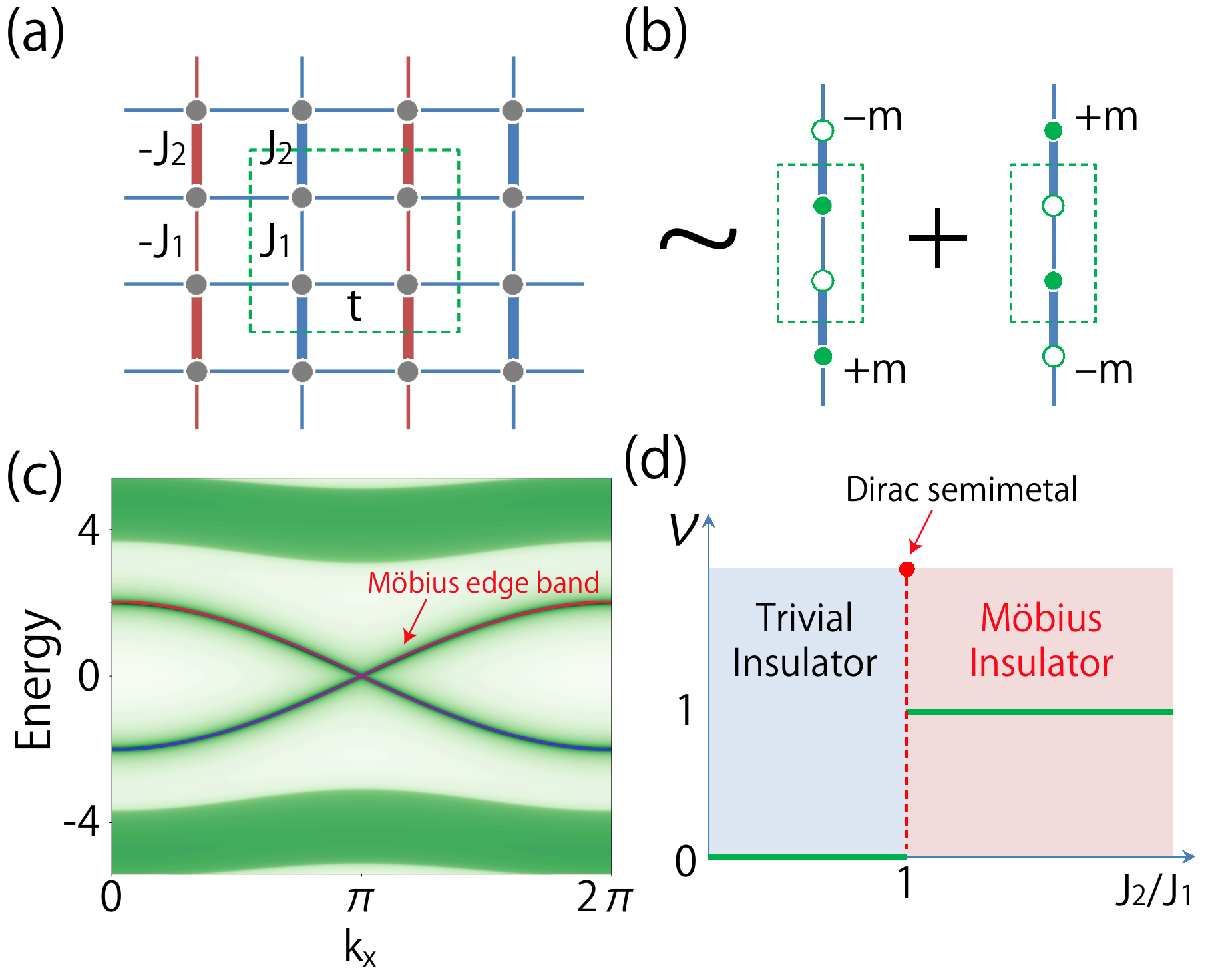}
	\caption{(a) illustrates the dimerization along $y$, which breaks the $\mathsf{L}_y$ symmetry. The hopping amplitudes are indicated in the figure. (b) For each $k_x$, the model can be mapped to two copies of the SSH model with a $k_x$-dependent mass term. The symmetry breaking transforms the Dirac semimetal into a M\"{o}bius topological insulator, with M\"{o}bius-twist edge bands for any edge along $x$. (c) shows the calculated edge spectrum for such an edge. Here, we take $t=1$, $J_1=1$, and $J_2=4$. (d) Phase diagram and $\Z_2$ invariant $\nu$ versus $J_2/J_1$. \label{fig:MTI}}
\end{figure}

One simplest way to achieve the desired symmetry breaking is through the dimerization along one direction. For instance, let's take dimerization along $y$ [see Fig.~\ref{fig:MTI}(a)]. Then, the symmetry-constrained lattice model can be written as
\begin{equation}\label{HH}
\begin{split}
\H(\bm k)=&t(1+\cos k_x)\Gamma^1+t\sin k_x\Gamma^2 \\ &+(J_1+J_2\cos k_y)\Gamma^3+J_2\sin k_y \Gamma^4,
\end{split}
\end{equation}
where $\Gamma^1=\tau_0\otimes\sigma_1$, $\Gamma^2=\tau_0\otimes\sigma_2$, $\Gamma^3=\tau_1\otimes\sigma_3$, $\Gamma^4=\tau_2\otimes\sigma_3$, and the real hopping amplitudes ($t, J_1, J_2>0$) are indicated in Fig.~\ref{fig:MTI}(a).
The dimerization corresponds to $J_1\neq J_2$, which breaks the primitive translation $\mathsf{L}_y$. It follows that the original Dirac point is destroyed, and the system becomes an insulator.
Meanwhile, $\mathsf{L}_x$ is still preserved.  In addition, the model has a sublattice symmetry $S$, which anti-commutes with $\mathsf{L}_x$,
\begin{equation}
\{\hat{S},\hat{\mathsf{L}}_x\}=0.
\end{equation}
As we show below, the two symmetries
lead to a $\Z_2$ classification of the resulting insulator phase.

Since
$
[{\mathsf{L}}_x,\H(\bm k)]=0,
$
we can perform a unitary transformation $U(k_x)$, so that $\hat{\mathsf{L}}_x$ is diagonalized as
\begin{equation}
U\hat{\mathsf{L}}_xU^\dagger=-e^{\i k_x/2}\tau_3\otimes\sigma_0,
\end{equation}
and the Hamiltonian is block diagonalized as
\begin{equation}\label{DD}
U\H U^\dagger=\left[\begin{matrix}
h_1(\bm k) & 0\\
0 & h_2(\bm k)
\end{matrix}\right].
\end{equation}
For the explicit expression of $U(k_x)$, see the SM~\cite{Supp}. The sublattice symmetry is transformed as
\begin{equation}
U\hat{S} U^\dagger=-\tau_1\otimes\sigma_3.
\end{equation}
Hence, sublattice symmetry requires that
\begin{equation}\label{SS}
\sigma_3 h_{1}(\bm k)\sigma_3=-h_2(\bm k).
\end{equation}
It is important to note that due to the projective nature of ${\mathsf{L}}_x$,
$U$ is not periodic in $k_x$. Specifically, a unit reciprocal translation gives
\begin{equation}\label{U-period}
U(k_x+2\pi)=U(k_x)V
\end{equation}
with $V=-\i \tau_1\otimes\sigma_0$~\cite{Supp}.
Consequently, $U\H U^\dagger$ is also not periodic in $k_x$, but satisfies the following relation
\begin{equation}\label{kx-period}
\sigma_3 h_{1,2}(k_x,k_y) \sigma_3=-h_{1,2}(k_x+2\pi,k_y).
\end{equation}

Explicitly, for our model in (\ref{HH}),
\begin{equation}
h_{1,2}(\bm k) =\left[\begin{matrix}
0 & q^*(k_y)\\
q(k_y) & 0
\end{matrix}\right]\pm m(k_x)\sigma_3,
\end{equation}
where $q(\bm k)=J_1+J_2e^{-\i k_y}$ and $m(k_x)=2t\cos (k_x/2)$. Interestingly, the first term is nothing but the standard Su-Schrieffer-Heeger (SSH) model~\cite{Su1980}. The second term is a mass term, depending only on $k_x$ [see Fig.~\ref{fig:MTI}(b)]. Varying $k_x$ from $0$ to $2\pi$, the mass term monotonically decreases from positive to negative, crossing zero at $k_x=\pi$. At $k_x=\pi$, the system is exactly two copies of SSH model. It is well known that SSH model is nontrivial (trivial) for $J_1<J_2$ ($J_1>J_2$), with (without) a zero-mode at each end. Thus, in the nontrivial phase, there is a pair of zero-modes at $k_x=\pi$ for an edge perpendicular to $y$. Deviating from $k_x=\pi$, the mass term shifts the edge modes away from zero energy. The pair at a given edge are shifted in opposite direction because of (\ref{SS}), forming edge bands crossing at $k_x=\pi$. Furthermore, due to (\ref{kx-period}), the edge bands are connected to each other, forming a M\"{o}bius twist over the edge BZ, as shown in Fig.~\ref{fig:MTI}(c). This peculiar connectivity is enforced by ${\mathsf{L}}_x$. Note that the two edge bands have opposite ${\mathsf{L}}_x$ eigenvalues $\pm e^{ik_x/2}$ which have a period of $4\pi$ and are inverted after wrapping around the BZ once, so M\"{o}bius-twist edge bands must exist at any ${\mathsf{L}}_x$-invariant edge for the nontrivial phase.

%

The above analysis is based on a mapping to the SSH model. Below, we present a $\Z_2$ topological invariant, which applies to general gapped systems preserving ${\mathsf{L}}_x$ and $S$ symmetries, not limited to the model (\ref{HH}). As ${\mathsf{L}}_x$ is preserved, the system Hamiltonian $\H(\bm k)$ can always be block-diagonalized into the two eigen-spaces of ${\mathsf{L}}_x$, as in Eq.~(\ref{DD}). The two eigen-spaces are connected by $S$, so we only need to focus on one of them, say $h_1(\bm k)$. The $\Z_2$ invariant is defined from the valence bands of $h_1$, given by
\begin{equation}
\nu=\frac{1}{2\pi}\int_{[0,2\pi]\times S^1} d^2k~ \mathcal{F}+\frac{1}{\pi}\gamma(0)\mod 2,
\end{equation}
where $\gamma(k_x)=\oint dk_y \mathcal{A}_y$ is the Berry phase for the 1D subsystem $h_1(k_x,k_y)$ with fixed $k_x$,  $
\mathcal{A}(\bm k)=\sum_{n}\langle\psi^-_n|\i\nabla_{\bm k}|\psi^-_n \rangle $ is the Berry connection for the valence bands of $h_1$, $\mathcal{F}=(\nabla_{\bm k}\times \mathcal{A})_z$ is the corresponding Berry curvature, the integration region of the first term is specified as $
[0,2\pi]\times S^1$ to emphasize that the Hamiltonian (hence the Berry curvature) is not periodic in $k_x$.
A nontrivial $\nu$ indicates a M\"{o}bius topological insulator, with M\"{o}bius-twist edge bands at any ${\mathsf{L}}_x$-invariant edge.
Applying the formula to (\ref{HH}), one finds that $\nu=1$ for $J_1<J_2$, and is trivial otherwise [see Fig.~\ref{fig:MTI}(d)], consistent with our previous analysis. More detailed derivations for the valence eigenstates and explanations for the topological invariant can be found in the SM~\cite{Supp}.


{\color{blue}\textit{Discussion}.} We have demonstrated that PRSG generates a new arena for exploring novel physics. Originated from the modifications of the fundamental algebraic structure, PRSG can lead to a twofold degenerate band structure, where translation operators form a topologically nontrivial spinor structure. In this work, we discussed two resulting topological phases, while many more are waiting to be explored.

We have seen that some PRSG symmetries possess features analogous to nonsymmorphic symmetries. Hence, the resulting topological phase, e.g., the M\"{o}bius topological insulator, may also find a counterpart protected by certain nonsymmorphic symmetry~\cite{Shiozaki2015,Zhao2016,Chang2017e,Zhang2020,Wieder1,Wieder2}. This analogy can be generalized into three dimensions.
However, there are also important differences. For instance, as a translation, ${\mathsf{L}}_x$ does not change $\bm k$, whereas nonsymmorphic operations typically do. Therefore, while ${\mathsf{L}}_x$ can enforce twist and crossing of a pair of bands along any axes along the translation direction in the BZ, non-symmorphic operator can only do it along invariant axes.


\begin{acknowledgments}
	{\color{blue}\textit{Acknowledgments.}} This work is supported by the NSFC (Grant No.~11874201), the Fundamental Research Funds for the Central Universities (Grant No.~0204/14380119), the Singapore MOE AcRF Tier 2 (MOE2019-T2-1-001), and the AIQ foundation of Nanjing University.
\end{acknowledgments}

\bibliographystyle{apsrev4-1}
\bibliography{Z2_Symm_REF}

\begin{thebibliography}{50}%
\makeatletter
\providecommand \@ifxundefined [1]{%
 \@ifx{#1\undefined}
}%
\providecommand \@ifnum [1]{%
 \ifnum #1\expandafter \@firstoftwo
 \else \expandafter \@secondoftwo
 \fi
}%
\providecommand \@ifx [1]{%
 \ifx #1\expandafter \@firstoftwo
 \else \expandafter \@secondoftwo
 \fi
}%
\providecommand \natexlab [1]{#1}%
\providecommand \enquote  [1]{``#1''}%
\providecommand \bibnamefont  [1]{#1}%
\providecommand \bibfnamefont [1]{#1}%
\providecommand \citenamefont [1]{#1}%
\providecommand \href@noop [0]{\@secondoftwo}%
\providecommand \href [0]{\begingroup \@sanitize@url \@href}%
\providecommand \@href[1]{\@@startlink{#1}\@@href}%
\providecommand \@@href[1]{\endgroup#1\@@endlink}%
\providecommand \@sanitize@url [0]{\catcode `\\12\catcode `\$12\catcode
  `\&12\catcode `\#12\catcode `\^12\catcode `\_12\catcode `\%12\relax}%
\providecommand \@@startlink[1]{}%
\providecommand \@@endlink[0]{}%
\providecommand \url  [0]{\begingroup\@sanitize@url \@url }%
\providecommand \@url [1]{\endgroup\@href {#1}{\urlprefix }}%
\providecommand \urlprefix  [0]{URL }%
\providecommand \Eprint [0]{\href }%
\providecommand \doibase [0]{http://dx.doi.org/}%
\providecommand \selectlanguage [0]{\@gobble}%
\providecommand \bibinfo  [0]{\@secondoftwo}%
\providecommand \bibfield  [0]{\@secondoftwo}%
\providecommand \translation [1]{[#1]}%
\providecommand \BibitemOpen [0]{}%
\providecommand \bibitemStop [0]{}%
\providecommand \bibitemNoStop [0]{.\EOS\space}%
\providecommand \EOS [0]{\spacefactor3000\relax}%
\providecommand \BibitemShut  [1]{\csname bibitem#1\endcsname}%
\let\auto@bib@innerbib\@empty
\bibitem [{\citenamefont {Klitzing}\ \emph {et~al.}(1980)\citenamefont
  {Klitzing}, \citenamefont {Dorda},\ and\ \citenamefont
  {Pepper}}]{Klitzing1980}%
  \BibitemOpen
  \bibfield  {author} {\bibinfo {author} {\bibfnamefont {K.~v.}\ \bibnamefont
  {Klitzing}}, \bibinfo {author} {\bibfnamefont {G.}~\bibnamefont {Dorda}}, \
  and\ \bibinfo {author} {\bibfnamefont {M.}~\bibnamefont {Pepper}},\ }\href
  {\doibase 10.1103/PhysRevLett.45.494} {\bibfield  {journal} {\bibinfo
  {journal} {Phys. Rev. Lett.}\ }\textbf {\bibinfo {volume} {45}},\ \bibinfo
  {pages} {494} (\bibinfo {year} {1980})}\BibitemShut {NoStop}%
\bibitem [{\citenamefont {Thouless}\ \emph {et~al.}(1982)\citenamefont
  {Thouless}, \citenamefont {Kohmoto}, \citenamefont {Nightingale},\ and\
  \citenamefont {den Nijs}}]{Thouless1982}%
  \BibitemOpen
  \bibfield  {author} {\bibinfo {author} {\bibfnamefont {D.~J.}\ \bibnamefont
  {Thouless}}, \bibinfo {author} {\bibfnamefont {M.}~\bibnamefont {Kohmoto}},
  \bibinfo {author} {\bibfnamefont {M.~P.}\ \bibnamefont {Nightingale}}, \ and\
  \bibinfo {author} {\bibfnamefont {M.}~\bibnamefont {den Nijs}},\ }\href
  {\doibase 10.1103/PhysRevLett.49.405} {\bibfield  {journal} {\bibinfo
  {journal} {Phys. Rev. Lett.}\ }\textbf {\bibinfo {volume} {49}},\ \bibinfo
  {pages} {405} (\bibinfo {year} {1982})}\BibitemShut {NoStop}%
\bibitem [{\citenamefont {Haldane}(1988)}]{Haldane1988}%
  \BibitemOpen
  \bibfield  {author} {\bibinfo {author} {\bibfnamefont {F.~D.~M.}\
  \bibnamefont {Haldane}},\ }\href {\doibase 10.1103/PhysRevLett.61.2015}
  {\bibfield  {journal} {\bibinfo  {journal} {Phys. Rev. Lett.}\ }\textbf
  {\bibinfo {volume} {61}},\ \bibinfo {pages} {2015} (\bibinfo {year}
  {1988})}\BibitemShut {NoStop}%
\bibitem [{\citenamefont {Chiu}\ \emph {et~al.}(2016)\citenamefont {Chiu},
  \citenamefont {Teo}, \citenamefont {Schnyder},\ and\ \citenamefont
  {Ryu}}]{Chiu2016}%
  \BibitemOpen
  \bibfield  {author} {\bibinfo {author} {\bibfnamefont {C.-K.}\ \bibnamefont
  {Chiu}}, \bibinfo {author} {\bibfnamefont {J.~C.~Y.}\ \bibnamefont {Teo}},
  \bibinfo {author} {\bibfnamefont {A.~P.}\ \bibnamefont {Schnyder}}, \ and\
  \bibinfo {author} {\bibfnamefont {S.}~\bibnamefont {Ryu}},\ }\href {\doibase
  10.1103/RevModPhys.88.035005} {\bibfield  {journal} {\bibinfo  {journal}
  {Rev. Mod. Phys.}\ }\textbf {\bibinfo {volume} {88}},\ \bibinfo {pages}
  {035005} (\bibinfo {year} {2016})}\BibitemShut {NoStop}%
\bibitem [{\citenamefont {Kane}\ and\ \citenamefont {Mele}(2005)}]{Kane2005a}%
  \BibitemOpen
  \bibfield  {author} {\bibinfo {author} {\bibfnamefont {C.~L.}\ \bibnamefont
  {Kane}}\ and\ \bibinfo {author} {\bibfnamefont {E.~J.}\ \bibnamefont
  {Mele}},\ }\href {\doibase 10.1103/PhysRevLett.95.146802} {\bibfield
  {journal} {\bibinfo  {journal} {Phys. Rev. Lett.}\ }\textbf {\bibinfo
  {volume} {95}},\ \bibinfo {pages} {146802} (\bibinfo {year}
  {2005})}\BibitemShut {NoStop}%
\bibitem [{\citenamefont {Schnyder}\ \emph {et~al.}(2008)\citenamefont
  {Schnyder}, \citenamefont {Ryu}, \citenamefont {Furusaki},\ and\
  \citenamefont {Ludwig}}]{Schnyder2008}%
  \BibitemOpen
  \bibfield  {author} {\bibinfo {author} {\bibfnamefont {A.~P.}\ \bibnamefont
  {Schnyder}}, \bibinfo {author} {\bibfnamefont {S.}~\bibnamefont {Ryu}},
  \bibinfo {author} {\bibfnamefont {A.}~\bibnamefont {Furusaki}}, \ and\
  \bibinfo {author} {\bibfnamefont {A.~W.~W.}\ \bibnamefont {Ludwig}},\ }\href
  {\doibase 10.1103/PhysRevB.78.195125} {\bibfield  {journal} {\bibinfo
  {journal} {Phys. Rev. B}\ }\textbf {\bibinfo {volume} {78}},\ \bibinfo
  {pages} {195125} (\bibinfo {year} {2008})}\BibitemShut {NoStop}%
\bibitem [{\citenamefont {Kitaev}(2009)}]{Kitaev2009}%
  \BibitemOpen
  \bibfield  {author} {\bibinfo {author} {\bibfnamefont {A.}~\bibnamefont
  {Kitaev}},\ }\href {\doibase http://dx.doi.org/10.1063/1.3149495} {\bibfield
  {journal} {\bibinfo  {journal} {AIP Conference Proceedings}\ }\textbf
  {\bibinfo {volume} {1134}},\ \bibinfo {pages} {22} (\bibinfo {year}
  {2009})}\BibitemShut {NoStop}%
\bibitem [{\citenamefont {Zhao}\ and\ \citenamefont {Wang}(2013)}]{Zhao2013b}%
  \BibitemOpen
  \bibfield  {author} {\bibinfo {author} {\bibfnamefont {Y.~X.}\ \bibnamefont
  {Zhao}}\ and\ \bibinfo {author} {\bibfnamefont {Z.~D.}\ \bibnamefont
  {Wang}},\ }\href {\doibase 10.1103/PhysRevLett.110.240404} {\bibfield
  {journal} {\bibinfo  {journal} {Phys. Rev. Lett.}\ }\textbf {\bibinfo
  {volume} {110}},\ \bibinfo {pages} {240404} (\bibinfo {year}
  {2013})}\BibitemShut {NoStop}%
\bibitem [{\citenamefont {Slager}\ \emph {et~al.}(2013)\citenamefont {Slager},
  \citenamefont {Mesaros}, \citenamefont {Juričić},\ and\ \citenamefont
  {Zaanen}}]{Slager2013}%
  \BibitemOpen
  \bibfield  {author} {\bibinfo {author} {\bibfnamefont {R.-J.}\ \bibnamefont
  {Slager}}, \bibinfo {author} {\bibfnamefont {A.}~\bibnamefont {Mesaros}},
  \bibinfo {author} {\bibfnamefont {V.}~\bibnamefont {Juričić}}, \ and\
  \bibinfo {author} {\bibfnamefont {J.}~\bibnamefont {Zaanen}},\ }\href
  {https://doi.org/10.1038/nphys2513} {\bibfield  {journal} {\bibinfo
  {journal} {Nature Physics}\ }\textbf {\bibinfo {volume} {9}},\ \bibinfo
  {pages} {98} (\bibinfo {year} {2013})}\BibitemShut {NoStop}%
\bibitem [{\citenamefont {Zhao}\ \emph {et~al.}(2016)\citenamefont {Zhao},
  \citenamefont {Schnyder},\ and\ \citenamefont {Wang}}]{Zhao2016a}%
  \BibitemOpen
  \bibfield  {author} {\bibinfo {author} {\bibfnamefont {Y.~X.}\ \bibnamefont
  {Zhao}}, \bibinfo {author} {\bibfnamefont {A.~P.}\ \bibnamefont {Schnyder}},
  \ and\ \bibinfo {author} {\bibfnamefont {Z.~D.}\ \bibnamefont {Wang}},\
  }\href {\doibase 10.1103/PhysRevLett.116.156402} {\bibfield  {journal}
  {\bibinfo  {journal} {Phys. Rev. Lett.}\ }\textbf {\bibinfo {volume} {116}},\
  \bibinfo {pages} {156402} (\bibinfo {year} {2016})}\BibitemShut {NoStop}%
\bibitem [{\citenamefont {Shiozaki}\ \emph {et~al.}(2016)\citenamefont
  {Shiozaki}, \citenamefont {Sato},\ and\ \citenamefont {Gomi}}]{Shiozaki2016}%
  \BibitemOpen
  \bibfield  {author} {\bibinfo {author} {\bibfnamefont {K.}~\bibnamefont
  {Shiozaki}}, \bibinfo {author} {\bibfnamefont {M.}~\bibnamefont {Sato}}, \
  and\ \bibinfo {author} {\bibfnamefont {K.}~\bibnamefont {Gomi}},\ }\href
  {\doibase 10.1103/PhysRevB.93.195413} {\bibfield  {journal} {\bibinfo
  {journal} {Phys. Rev. B}\ }\textbf {\bibinfo {volume} {93}},\ \bibinfo
  {pages} {195413} (\bibinfo {year} {2016})}\BibitemShut {NoStop}%
\bibitem [{\citenamefont {Bradlyn}\ \emph {et~al.}(2017)\citenamefont
  {Bradlyn}, \citenamefont {Elcoro}, \citenamefont {Cano}, \citenamefont
  {Vergniory}, \citenamefont {Wang}, \citenamefont {Felser}, \citenamefont
  {Aroyo},\ and\ \citenamefont {Bernevig}}]{Bradlyn2017}%
  \BibitemOpen
  \bibfield  {author} {\bibinfo {author} {\bibfnamefont {B.}~\bibnamefont
  {Bradlyn}}, \bibinfo {author} {\bibfnamefont {L.}~\bibnamefont {Elcoro}},
  \bibinfo {author} {\bibfnamefont {J.}~\bibnamefont {Cano}}, \bibinfo {author}
  {\bibfnamefont {M.~G.}\ \bibnamefont {Vergniory}}, \bibinfo {author}
  {\bibfnamefont {Z.}~\bibnamefont {Wang}}, \bibinfo {author} {\bibfnamefont
  {C.}~\bibnamefont {Felser}}, \bibinfo {author} {\bibfnamefont {M.~I.}\
  \bibnamefont {Aroyo}}, \ and\ \bibinfo {author} {\bibfnamefont {B.~A.}\
  \bibnamefont {Bernevig}},\ }\href {https://doi.org/10.1038/nature23268}
  {\bibfield  {journal} {\bibinfo  {journal} {Nature}\ }\textbf {\bibinfo
  {volume} {547}},\ \bibinfo {pages} {298} (\bibinfo {year}
  {2017})}\BibitemShut {NoStop}%
\bibitem [{\citenamefont {Tang}\ \emph
  {et~al.}(2019{\natexlab{a}})\citenamefont {Tang}, \citenamefont {Po},
  \citenamefont {Vishwanath},\ and\ \citenamefont {Wan}}]{Tang2019}%
  \BibitemOpen
  \bibfield  {author} {\bibinfo {author} {\bibfnamefont {F.}~\bibnamefont
  {Tang}}, \bibinfo {author} {\bibfnamefont {H.~C.}\ \bibnamefont {Po}},
  \bibinfo {author} {\bibfnamefont {A.}~\bibnamefont {Vishwanath}}, \ and\
  \bibinfo {author} {\bibfnamefont {X.}~\bibnamefont {Wan}},\ }\href@noop {}
  {\bibfield  {journal} {\bibinfo  {journal} {Nature Physics}\ }\textbf
  {\bibinfo {volume} {15}},\ \bibinfo {pages} {470} (\bibinfo {year}
  {2019}{\natexlab{a}})}\BibitemShut {NoStop}%
\bibitem [{\citenamefont {Zhang}\ \emph {et~al.}(2019)\citenamefont {Zhang},
  \citenamefont {Jiang}, \citenamefont {Song}, \citenamefont {Huang},
  \citenamefont {He}, \citenamefont {Fang}, \citenamefont {Weng},\ and\
  \citenamefont {Fang}}]{Zhang2019}%
  \BibitemOpen
  \bibfield  {author} {\bibinfo {author} {\bibfnamefont {T.}~\bibnamefont
  {Zhang}}, \bibinfo {author} {\bibfnamefont {Y.}~\bibnamefont {Jiang}},
  \bibinfo {author} {\bibfnamefont {Z.}~\bibnamefont {Song}}, \bibinfo {author}
  {\bibfnamefont {H.}~\bibnamefont {Huang}}, \bibinfo {author} {\bibfnamefont
  {Y.}~\bibnamefont {He}}, \bibinfo {author} {\bibfnamefont {Z.}~\bibnamefont
  {Fang}}, \bibinfo {author} {\bibfnamefont {H.}~\bibnamefont {Weng}}, \ and\
  \bibinfo {author} {\bibfnamefont {C.}~\bibnamefont {Fang}},\ }\href
  {https://doi.org/10.1038/s41586-019-0944-6} {\bibfield  {journal} {\bibinfo
  {journal} {Nature}\ }\textbf {\bibinfo {volume} {566}},\ \bibinfo {pages}
  {475} (\bibinfo {year} {2019})}\BibitemShut {NoStop}%
\bibitem [{\citenamefont {Vergniory}\ \emph {et~al.}(2019)\citenamefont
  {Vergniory}, \citenamefont {Elcoro}, \citenamefont {Felser}, \citenamefont
  {Regnault}, \citenamefont {Bernevig},\ and\ \citenamefont
  {Wang}}]{Vergniory2019}%
  \BibitemOpen
  \bibfield  {author} {\bibinfo {author} {\bibfnamefont {M.~G.}\ \bibnamefont
  {Vergniory}}, \bibinfo {author} {\bibfnamefont {L.}~\bibnamefont {Elcoro}},
  \bibinfo {author} {\bibfnamefont {C.}~\bibnamefont {Felser}}, \bibinfo
  {author} {\bibfnamefont {N.}~\bibnamefont {Regnault}}, \bibinfo {author}
  {\bibfnamefont {B.~A.}\ \bibnamefont {Bernevig}}, \ and\ \bibinfo {author}
  {\bibfnamefont {Z.}~\bibnamefont {Wang}},\ }\href
  {https://doi.org/10.1038/s41586-019-0954-4} {\bibfield  {journal} {\bibinfo
  {journal} {Nature}\ }\textbf {\bibinfo {volume} {566}},\ \bibinfo {pages}
  {480} (\bibinfo {year} {2019})}\BibitemShut {NoStop}%
\bibitem [{\citenamefont {Tang}\ \emph
  {et~al.}(2019{\natexlab{b}})\citenamefont {Tang}, \citenamefont {Po},
  \citenamefont {Vishwanath},\ and\ \citenamefont {Wan}}]{Tang2019a}%
  \BibitemOpen
  \bibfield  {author} {\bibinfo {author} {\bibfnamefont {F.}~\bibnamefont
  {Tang}}, \bibinfo {author} {\bibfnamefont {H.~C.}\ \bibnamefont {Po}},
  \bibinfo {author} {\bibfnamefont {A.}~\bibnamefont {Vishwanath}}, \ and\
  \bibinfo {author} {\bibfnamefont {X.}~\bibnamefont {Wan}},\ }\href
  {https://doi.org/10.1038/s41586-019-0937-5} {\bibfield  {journal} {\bibinfo
  {journal} {Nature}\ }\textbf {\bibinfo {volume} {566}},\ \bibinfo {pages}
  {486} (\bibinfo {year} {2019}{\natexlab{b}})}\BibitemShut {NoStop}%
\bibitem [{\citenamefont {Moore}(2020)}]{moore_Abstract_Group}%
  \BibitemOpen
  \bibfield  {author} {\bibinfo {author} {\bibfnamefont {G.~W.}\ \bibnamefont
  {Moore}},\ }\href@noop {} {\enquote {\bibinfo {title} {Lecture notes:
  Abstract group theory},}\ } (\bibinfo {year} {2020})\BibitemShut {NoStop}%
\bibitem [{\citenamefont {Hasan}\ and\ \citenamefont {Kane}(2010)}]{Hasan2010}%
  \BibitemOpen
  \bibfield  {author} {\bibinfo {author} {\bibfnamefont {M.~Z.}\ \bibnamefont
  {Hasan}}\ and\ \bibinfo {author} {\bibfnamefont {C.~L.}\ \bibnamefont
  {Kane}},\ }\href {\doibase 10.1103/RevModPhys.82.3045} {\bibfield  {journal}
  {\bibinfo  {journal} {Rev. Mod. Phys.}\ }\textbf {\bibinfo {volume} {82}},\
  \bibinfo {pages} {3045} (\bibinfo {year} {2010})}\BibitemShut {NoStop}%
\bibitem [{\citenamefont {Qi}\ and\ \citenamefont {Zhang}(2011)}]{Qi2011}%
  \BibitemOpen
  \bibfield  {author} {\bibinfo {author} {\bibfnamefont {X.-L.}\ \bibnamefont
  {Qi}}\ and\ \bibinfo {author} {\bibfnamefont {S.-C.}\ \bibnamefont {Zhang}},\
  }\href {\doibase 10.1103/RevModPhys.83.1057} {\bibfield  {journal} {\bibinfo
  {journal} {Rev. Mod. Phys.}\ }\textbf {\bibinfo {volume} {83}},\ \bibinfo
  {pages} {1057} (\bibinfo {year} {2011})}\BibitemShut {NoStop}%
\bibitem [{\citenamefont {Bais}\ \emph {et~al.}(1992)\citenamefont {Bais},
  \citenamefont {van Driel},\ and\ \citenamefont
  {de~Wild~Propitius}}]{Discrete_Gauge}%
  \BibitemOpen
  \bibfield  {author} {\bibinfo {author} {\bibfnamefont {F.~A.}\ \bibnamefont
  {Bais}}, \bibinfo {author} {\bibfnamefont {P.}~\bibnamefont {van Driel}}, \
  and\ \bibinfo {author} {\bibfnamefont {M.}~\bibnamefont
  {de~Wild~Propitius}},\ }\href {\doibase
  https://doi.org/10.1016/0370-2693(92)90773-W} {\bibfield  {journal} {\bibinfo
   {journal} {Physics Letters B}\ }\textbf {\bibinfo {volume} {280}},\ \bibinfo
  {pages} {63 } (\bibinfo {year} {1992})}\BibitemShut {NoStop}%
\bibitem [{\citenamefont {Hansson}\ \emph {et~al.}(2004)\citenamefont
  {Hansson}, \citenamefont {Oganesyan},\ and\ \citenamefont
  {Sondhi}}]{Superconductor_Z2}%
  \BibitemOpen
  \bibfield  {author} {\bibinfo {author} {\bibfnamefont {T.}~\bibnamefont
  {Hansson}}, \bibinfo {author} {\bibfnamefont {V.}~\bibnamefont {Oganesyan}},
  \ and\ \bibinfo {author} {\bibfnamefont {S.}~\bibnamefont {Sondhi}},\ }\href
  {\doibase https://doi.org/10.1016/j.aop.2004.05.006} {\bibfield  {journal}
  {\bibinfo  {journal} {Annals of Physics}\ }\textbf {\bibinfo {volume}
  {313}},\ \bibinfo {pages} {497 } (\bibinfo {year} {2004})}\BibitemShut
  {NoStop}%
\bibitem [{\citenamefont {Wen}(2017)}]{Xiao-Gang_RMP}%
  \BibitemOpen
  \bibfield  {author} {\bibinfo {author} {\bibfnamefont {X.-G.}\ \bibnamefont
  {Wen}},\ }\href {\doibase 10.1103/RevModPhys.89.041004} {\bibfield  {journal}
  {\bibinfo  {journal} {Rev. Mod. Phys.}\ }\textbf {\bibinfo {volume} {89}},\
  \bibinfo {pages} {041004} (\bibinfo {year} {2017})}\BibitemShut {NoStop}%
\bibitem [{\citenamefont {Kitaev}(2006)}]{Kitaev2006}%
  \BibitemOpen
  \bibfield  {author} {\bibinfo {author} {\bibfnamefont {A.}~\bibnamefont
  {Kitaev}},\ }\bibfield  {booktitle} {\emph {\bibinfo {booktitle} {January
  Special Issue}},\ }\href
  {http://www.sciencedirect.com/science/article/pii/S0003491605002381}
  {\bibfield  {journal} {\bibinfo  {journal} {Annals of Physics}\ }\textbf
  {\bibinfo {volume} {321}},\ \bibinfo {pages} {2} (\bibinfo {year}
  {2006})}\BibitemShut {NoStop}%
\bibitem [{\citenamefont {Savary}\ and\ \citenamefont
  {Balents}(2016)}]{Savary2016}%
  \BibitemOpen
  \bibfield  {author} {\bibinfo {author} {\bibfnamefont {L.}~\bibnamefont
  {Savary}}\ and\ \bibinfo {author} {\bibfnamefont {L.}~\bibnamefont
  {Balents}},\ }\bibfield  {booktitle} {\emph {\bibinfo {booktitle} {Reports on
  Progress in Physics}},\ }\href
  {http://dx.doi.org/10.1088/0034-4885/80/1/016502} {\ \textbf {\bibinfo
  {volume} {80}},\ \bibinfo {pages} {016502} (\bibinfo {year}
  {2016})}\BibitemShut {NoStop}%
\bibitem [{\citenamefont {Zhao}\ \emph {et~al.}()\citenamefont {Zhao},
  \citenamefont {Lu},\ and\ \citenamefont {Yang}}]{Zhao2020}%
  \BibitemOpen
  \bibfield  {author} {\bibinfo {author} {\bibfnamefont {Y.~X.}\ \bibnamefont
  {Zhao}}, \bibinfo {author} {\bibfnamefont {Y.}~\bibnamefont {Lu}}, \ and\
  \bibinfo {author} {\bibfnamefont {S.~A.}\ \bibnamefont {Yang}},\ }\href@noop
  {} {}\bibinfo {note} {ArXiv:2005.14500}\BibitemShut {NoStop}%
\bibitem [{\citenamefont {Lu}\ \emph {et~al.}(2014)\citenamefont {Lu},
  \citenamefont {Joannopoulos},\ and\ \citenamefont {Soljačić}}]{Lu2014}%
  \BibitemOpen
  \bibfield  {author} {\bibinfo {author} {\bibfnamefont {L.}~\bibnamefont
  {Lu}}, \bibinfo {author} {\bibfnamefont {J.~D.}\ \bibnamefont
  {Joannopoulos}}, \ and\ \bibinfo {author} {\bibfnamefont {M.}~\bibnamefont
  {Soljačić}},\ }\href {https://doi.org/10.1038/nphoton.2014.248} {\bibfield
  {journal} {\bibinfo  {journal} {Nature Photonics}\ }\textbf {\bibinfo
  {volume} {8}},\ \bibinfo {pages} {821} (\bibinfo {year} {2014})}\BibitemShut
  {NoStop}%
\bibitem [{\citenamefont {Yang}\ \emph {et~al.}(2015)\citenamefont {Yang},
  \citenamefont {Gao}, \citenamefont {Shi}, \citenamefont {Lin}, \citenamefont
  {Gao}, \citenamefont {Chong},\ and\ \citenamefont {Zhang}}]{Yang2015}%
  \BibitemOpen
  \bibfield  {author} {\bibinfo {author} {\bibfnamefont {Z.}~\bibnamefont
  {Yang}}, \bibinfo {author} {\bibfnamefont {F.}~\bibnamefont {Gao}}, \bibinfo
  {author} {\bibfnamefont {X.}~\bibnamefont {Shi}}, \bibinfo {author}
  {\bibfnamefont {X.}~\bibnamefont {Lin}}, \bibinfo {author} {\bibfnamefont
  {Z.}~\bibnamefont {Gao}}, \bibinfo {author} {\bibfnamefont {Y.}~\bibnamefont
  {Chong}}, \ and\ \bibinfo {author} {\bibfnamefont {B.}~\bibnamefont
  {Zhang}},\ }\href {\doibase 10.1103/PhysRevLett.114.114301} {\bibfield
  {journal} {\bibinfo  {journal} {Phys. Rev. Lett.}\ }\textbf {\bibinfo
  {volume} {114}},\ \bibinfo {pages} {114301} (\bibinfo {year}
  {2015})}\BibitemShut {NoStop}%
\bibitem [{\citenamefont {Mittal}\ \emph {et~al.}(2019)\citenamefont {Mittal},
  \citenamefont {Orre}, \citenamefont {Zhu}, \citenamefont {Gorlach},
  \citenamefont {Poddubny},\ and\ \citenamefont
  {Hafezi}}]{Photonic_Crystal_quadrupole}%
  \BibitemOpen
  \bibfield  {author} {\bibinfo {author} {\bibfnamefont {S.}~\bibnamefont
  {Mittal}}, \bibinfo {author} {\bibfnamefont {V.~V.}\ \bibnamefont {Orre}},
  \bibinfo {author} {\bibfnamefont {G.}~\bibnamefont {Zhu}}, \bibinfo {author}
  {\bibfnamefont {M.~A.}\ \bibnamefont {Gorlach}}, \bibinfo {author}
  {\bibfnamefont {A.}~\bibnamefont {Poddubny}}, \ and\ \bibinfo {author}
  {\bibfnamefont {M.}~\bibnamefont {Hafezi}},\ }\href {\doibase
  10.1038/s41566-019-0452-0} {\bibfield  {journal} {\bibinfo  {journal} {Nat.
  Photon.}\ }\textbf {\bibinfo {volume} {13}},\ \bibinfo {pages} {692–696}
  (\bibinfo {year} {2019})}\BibitemShut {NoStop}%
\bibitem [{\citenamefont {Xue}\ \emph {et~al.}(2020)\citenamefont {Xue},
  \citenamefont {Ge}, \citenamefont {Sun}, \citenamefont {Wang}, \citenamefont
  {Jia}, \citenamefont {Guan}, \citenamefont {Yuan}, \citenamefont {Chong},\
  and\ \citenamefont {Zhang}}]{Acoustic_Crystal}%
  \BibitemOpen
  \bibfield  {author} {\bibinfo {author} {\bibfnamefont {H.}~\bibnamefont
  {Xue}}, \bibinfo {author} {\bibfnamefont {Y.}~\bibnamefont {Ge}}, \bibinfo
  {author} {\bibfnamefont {H.-X.}\ \bibnamefont {Sun}}, \bibinfo {author}
  {\bibfnamefont {Q.}~\bibnamefont {Wang}}, \bibinfo {author} {\bibfnamefont
  {D.}~\bibnamefont {Jia}}, \bibinfo {author} {\bibfnamefont {Y.-J.}\
  \bibnamefont {Guan}}, \bibinfo {author} {\bibfnamefont {S.-Q.}\ \bibnamefont
  {Yuan}}, \bibinfo {author} {\bibfnamefont {Y.}~\bibnamefont {Chong}}, \ and\
  \bibinfo {author} {\bibfnamefont {B.}~\bibnamefont {Zhang}},\ }\href@noop {}
  {\bibfield  {journal} {\bibinfo  {journal} {Nat. Commun.}\ }\textbf {\bibinfo
  {volume} {11}},\ \bibinfo {pages} {2442} (\bibinfo {year}
  {2020})}\BibitemShut {NoStop}%
\bibitem [{\citenamefont {Imhof}\ \emph {et~al.}(2018)\citenamefont {Imhof},
  \citenamefont {Berger}, \citenamefont {Bayer}, \citenamefont {Brehm},
  \citenamefont {Molenkamp}, \citenamefont {Kiessling}, \citenamefont
  {Schindler}, \citenamefont {Lee}, \citenamefont {Greiter}, \citenamefont
  {Neupert},\ and\ \citenamefont {Thomale}}]{Imhof2018}%
  \BibitemOpen
  \bibfield  {author} {\bibinfo {author} {\bibfnamefont {S.}~\bibnamefont
  {Imhof}}, \bibinfo {author} {\bibfnamefont {C.}~\bibnamefont {Berger}},
  \bibinfo {author} {\bibfnamefont {F.}~\bibnamefont {Bayer}}, \bibinfo
  {author} {\bibfnamefont {J.}~\bibnamefont {Brehm}}, \bibinfo {author}
  {\bibfnamefont {L.~W.}\ \bibnamefont {Molenkamp}}, \bibinfo {author}
  {\bibfnamefont {T.}~\bibnamefont {Kiessling}}, \bibinfo {author}
  {\bibfnamefont {F.}~\bibnamefont {Schindler}}, \bibinfo {author}
  {\bibfnamefont {C.~H.}\ \bibnamefont {Lee}}, \bibinfo {author} {\bibfnamefont
  {M.}~\bibnamefont {Greiter}}, \bibinfo {author} {\bibfnamefont
  {T.}~\bibnamefont {Neupert}}, \ and\ \bibinfo {author} {\bibfnamefont
  {R.}~\bibnamefont {Thomale}},\ }\href
  {https://doi.org/10.1038/s41567-018-0246-1} {\bibfield  {journal} {\bibinfo
  {journal} {Nature Physics}\ }\textbf {\bibinfo {volume} {14}},\ \bibinfo
  {pages} {925} (\bibinfo {year} {2018})}\BibitemShut {NoStop}%
\bibitem [{\citenamefont {Yu}\ \emph {et~al.}(2020)\citenamefont {Yu},
  \citenamefont {Zhao},\ and\ \citenamefont {Schnyder}}]{Yu2020}%
  \BibitemOpen
  \bibfield  {author} {\bibinfo {author} {\bibfnamefont {R.}~\bibnamefont
  {Yu}}, \bibinfo {author} {\bibfnamefont {Y.~X.}\ \bibnamefont {Zhao}}, \ and\
  \bibinfo {author} {\bibfnamefont {A.~P.}\ \bibnamefont {Schnyder}},\
  }\href@noop {} {\bibfield  {journal} {\bibinfo  {journal} {National Science
  Review}\ }\textbf {\bibinfo {volume} {7}},\ \bibinfo {pages} {1288} (\bibinfo
  {year} {2020})}\BibitemShut {NoStop}%
\bibitem [{\citenamefont {Huber}(2016)}]{Huber2016}%
  \BibitemOpen
  \bibfield  {author} {\bibinfo {author} {\bibfnamefont {S.~D.}\ \bibnamefont
  {Huber}},\ }\href {https://doi.org/10.1038/nphys3801} {\bibfield  {journal}
  {\bibinfo  {journal} {Nature Physics}\ }\textbf {\bibinfo {volume} {12}},\
  \bibinfo {pages} {621} (\bibinfo {year} {2016})}\BibitemShut {NoStop}%
\bibitem [{\citenamefont {Prodan}\ and\ \citenamefont
  {Prodan}(2009)}]{Prodan_Spring}%
  \BibitemOpen
  \bibfield  {author} {\bibinfo {author} {\bibfnamefont {E.}~\bibnamefont
  {Prodan}}\ and\ \bibinfo {author} {\bibfnamefont {C.}~\bibnamefont
  {Prodan}},\ }\href {\doibase 10.1103/PhysRevLett.103.248101} {\bibfield
  {journal} {\bibinfo  {journal} {Phys. Rev. Lett.}\ }\textbf {\bibinfo
  {volume} {103}},\ \bibinfo {pages} {248101} (\bibinfo {year}
  {2009})}\BibitemShut {NoStop}%
\bibitem [{Sup()}]{Supp}%
  \BibitemOpen
  \href@noop {} {}\bibinfo {note} {See the Supplemental Materials.}\BibitemShut
  {Stop}%
\bibitem [{\citenamefont {Lieb}(1994)}]{Lieb_Theorem}%
  \BibitemOpen
  \bibfield  {author} {\bibinfo {author} {\bibfnamefont {E.~H.}\ \bibnamefont
  {Lieb}},\ }\href {\doibase 10.1103/PhysRevLett.73.2158} {\bibfield  {journal}
  {\bibinfo  {journal} {Phys. Rev. Lett.}\ }\textbf {\bibinfo {volume} {73}},\
  \bibinfo {pages} {2158} (\bibinfo {year} {1994})}\BibitemShut {NoStop}%
\bibitem [{\citenamefont {Young}\ \emph {et~al.}(2012)\citenamefont {Young},
  \citenamefont {Zaheer}, \citenamefont {Teo}, \citenamefont {Kane},
  \citenamefont {Mele},\ and\ \citenamefont {Rappe}}]{Young2012}%
  \BibitemOpen
  \bibfield  {author} {\bibinfo {author} {\bibfnamefont {S.~M.}\ \bibnamefont
  {Young}}, \bibinfo {author} {\bibfnamefont {S.}~\bibnamefont {Zaheer}},
  \bibinfo {author} {\bibfnamefont {J.~C.~Y.}\ \bibnamefont {Teo}}, \bibinfo
  {author} {\bibfnamefont {C.~L.}\ \bibnamefont {Kane}}, \bibinfo {author}
  {\bibfnamefont {E.~J.}\ \bibnamefont {Mele}}, \ and\ \bibinfo {author}
  {\bibfnamefont {A.~M.}\ \bibnamefont {Rappe}},\ }\href {\doibase
  10.1103/PhysRevLett.108.140405} {\bibfield  {journal} {\bibinfo  {journal}
  {Phys. Rev. Lett.}\ }\textbf {\bibinfo {volume} {108}},\ \bibinfo {pages}
  {140405} (\bibinfo {year} {2012})}\BibitemShut {NoStop}%
\bibitem [{\citenamefont {Michel}\ and\ \citenamefont
  {Zak}(1999)}]{Zak-Nonsymmorphic}%
  \BibitemOpen
  \bibfield  {author} {\bibinfo {author} {\bibfnamefont {L.}~\bibnamefont
  {Michel}}\ and\ \bibinfo {author} {\bibfnamefont {J.}~\bibnamefont {Zak}},\
  }\href {\doibase 10.1103/PhysRevB.59.5998} {\bibfield  {journal} {\bibinfo
  {journal} {Phys. Rev. B}\ }\textbf {\bibinfo {volume} {59}},\ \bibinfo
  {pages} {5998} (\bibinfo {year} {1999})}\BibitemShut {NoStop}%
\bibitem [{\citenamefont {Zhao}\ and\ \citenamefont
  {Schnyder}(2016)}]{Zhao2016}%
  \BibitemOpen
  \bibfield  {author} {\bibinfo {author} {\bibfnamefont {Y.~X.}\ \bibnamefont
  {Zhao}}\ and\ \bibinfo {author} {\bibfnamefont {A.~P.}\ \bibnamefont
  {Schnyder}},\ }\href {\doibase 10.1103/PhysRevB.94.195109} {\bibfield
  {journal} {\bibinfo  {journal} {Phys. Rev. B}\ }\textbf {\bibinfo {volume}
  {94}},\ \bibinfo {pages} {195109} (\bibinfo {year} {2016})}\BibitemShut
  {NoStop}%
\bibitem [{\citenamefont {Young}\ and\ \citenamefont {Kane}(2015)}]{Young2015}%
  \BibitemOpen
  \bibfield  {author} {\bibinfo {author} {\bibfnamefont {S.~M.}\ \bibnamefont
  {Young}}\ and\ \bibinfo {author} {\bibfnamefont {C.~L.}\ \bibnamefont
  {Kane}},\ }\href {\doibase 10.1103/PhysRevLett.115.126803} {\bibfield
  {journal} {\bibinfo  {journal} {Phys. Rev. Lett.}\ }\textbf {\bibinfo
  {volume} {115}},\ \bibinfo {pages} {126803} (\bibinfo {year}
  {2015})}\BibitemShut {NoStop}%
\bibitem [{\citenamefont {Shiozaki}\ \emph {et~al.}(2015)\citenamefont
  {Shiozaki}, \citenamefont {Sato},\ and\ \citenamefont {Gomi}}]{Shiozaki2015}%
  \BibitemOpen
  \bibfield  {author} {\bibinfo {author} {\bibfnamefont {K.}~\bibnamefont
  {Shiozaki}}, \bibinfo {author} {\bibfnamefont {M.}~\bibnamefont {Sato}}, \
  and\ \bibinfo {author} {\bibfnamefont {K.}~\bibnamefont {Gomi}},\ }\href
  {\doibase 10.1103/PhysRevB.91.155120} {\bibfield  {journal} {\bibinfo
  {journal} {Phys. Rev. B}\ }\textbf {\bibinfo {volume} {91}},\ \bibinfo
  {pages} {155120} (\bibinfo {year} {2015})}\BibitemShut {NoStop}%
\bibitem [{\citenamefont {Wang}\ \emph {et~al.}(2016)\citenamefont {Wang},
  \citenamefont {Alexandradinata}, \citenamefont {Cava},\ and\ \citenamefont
  {Bernevig}}]{Wang2016a}%
  \BibitemOpen
  \bibfield  {author} {\bibinfo {author} {\bibfnamefont {Z.}~\bibnamefont
  {Wang}}, \bibinfo {author} {\bibfnamefont {A.}~\bibnamefont
  {Alexandradinata}}, \bibinfo {author} {\bibfnamefont {R.~J.}\ \bibnamefont
  {Cava}}, \ and\ \bibinfo {author} {\bibfnamefont {B.~A.}\ \bibnamefont
  {Bernevig}},\ }\href {http://dx.doi.org/10.1038/nature17410} {\bibfield
  {journal} {\bibinfo  {journal} {Nature}\ }\textbf {\bibinfo {volume} {532}},\
  \bibinfo {pages} {189} (\bibinfo {year} {2016})}\BibitemShut {NoStop}%
\bibitem [{\citenamefont {Watanabe}\ \emph {et~al.}(2016)\citenamefont
  {Watanabe}, \citenamefont {Po}, \citenamefont {Zaletel},\ and\ \citenamefont
  {Vishwanath}}]{Watanabe2016}%
  \BibitemOpen
  \bibfield  {author} {\bibinfo {author} {\bibfnamefont {H.}~\bibnamefont
  {Watanabe}}, \bibinfo {author} {\bibfnamefont {H.~C.}\ \bibnamefont {Po}},
  \bibinfo {author} {\bibfnamefont {M.~P.}\ \bibnamefont {Zaletel}}, \ and\
  \bibinfo {author} {\bibfnamefont {A.}~\bibnamefont {Vishwanath}},\ }\href
  {\doibase 10.1103/PhysRevLett.117.096404} {\bibfield  {journal} {\bibinfo
  {journal} {Phys. Rev. Lett.}\ }\textbf {\bibinfo {volume} {117}},\ \bibinfo
  {pages} {096404} (\bibinfo {year} {2016})}\BibitemShut {NoStop}%
\bibitem [{\citenamefont {Bzdusek}\ \emph {et~al.}(2016)\citenamefont
  {Bzdusek}, \citenamefont {Wu}, \citenamefont {Ruegg}, \citenamefont
  {Sigrist},\ and\ \citenamefont {Soluyanov}}]{Bzdusek2016}%
  \BibitemOpen
  \bibfield  {author} {\bibinfo {author} {\bibfnamefont {T.}~\bibnamefont
  {Bzdusek}}, \bibinfo {author} {\bibfnamefont {Q.}~\bibnamefont {Wu}},
  \bibinfo {author} {\bibfnamefont {A.}~\bibnamefont {Ruegg}}, \bibinfo
  {author} {\bibfnamefont {M.}~\bibnamefont {Sigrist}}, \ and\ \bibinfo
  {author} {\bibfnamefont {A.~A.}\ \bibnamefont {Soluyanov}},\ }\href
  {http://dx.doi.org/10.1038/nature19099} {\bibfield  {journal} {\bibinfo
  {journal} {Nature}\ }\textbf {\bibinfo {volume} {538}},\ \bibinfo {pages}
  {75} (\bibinfo {year} {2016})}\BibitemShut {NoStop}%
\bibitem [{\citenamefont {Chang}\ \emph {et~al.}(2017)\citenamefont {Chang},
  \citenamefont {Erten},\ and\ \citenamefont {Coleman}}]{Chang2017e}%
  \BibitemOpen
  \bibfield  {author} {\bibinfo {author} {\bibfnamefont {P.-Y.}\ \bibnamefont
  {Chang}}, \bibinfo {author} {\bibfnamefont {O.}~\bibnamefont {Erten}}, \ and\
  \bibinfo {author} {\bibfnamefont {P.}~\bibnamefont {Coleman}},\ }\href
  {https://doi.org/10.1038/nphys4092} {\bibfield  {journal} {\bibinfo
  {journal} {Nature Physics}\ }\textbf {\bibinfo {volume} {13}},\ \bibinfo
  {pages} {794} (\bibinfo {year} {2017})}\BibitemShut {NoStop}%
\bibitem [{\citenamefont {Wu}\ \emph {et~al.}(2018)\citenamefont {Wu},
  \citenamefont {Liu}, \citenamefont {Li}, \citenamefont {Zhong}, \citenamefont
  {Yu}, \citenamefont {Sheng}, \citenamefont {Zhao},\ and\ \citenamefont
  {Yang}}]{Wu2018c}%
  \BibitemOpen
  \bibfield  {author} {\bibinfo {author} {\bibfnamefont {W.}~\bibnamefont
  {Wu}}, \bibinfo {author} {\bibfnamefont {Y.}~\bibnamefont {Liu}}, \bibinfo
  {author} {\bibfnamefont {S.}~\bibnamefont {Li}}, \bibinfo {author}
  {\bibfnamefont {C.}~\bibnamefont {Zhong}}, \bibinfo {author} {\bibfnamefont
  {Z.-M.}\ \bibnamefont {Yu}}, \bibinfo {author} {\bibfnamefont {X.-L.}\
  \bibnamefont {Sheng}}, \bibinfo {author} {\bibfnamefont {Y.~X.}\ \bibnamefont
  {Zhao}}, \ and\ \bibinfo {author} {\bibfnamefont {S.~A.}\ \bibnamefont
  {Yang}},\ }\href {\doibase 10.1103/PhysRevB.97.115125} {\bibfield  {journal}
  {\bibinfo  {journal} {Phys. Rev. B}\ }\textbf {\bibinfo {volume} {97}},\
  \bibinfo {pages} {115125} (\bibinfo {year} {2018})}\BibitemShut {NoStop}%
\bibitem [{\citenamefont {Yu}\ \emph {et~al.}(2019)\citenamefont {Yu},
  \citenamefont {Wu}, \citenamefont {Zhao},\ and\ \citenamefont
  {Yang}}]{Yu2019b}%
  \BibitemOpen
  \bibfield  {author} {\bibinfo {author} {\bibfnamefont {Z.-M.}\ \bibnamefont
  {Yu}}, \bibinfo {author} {\bibfnamefont {W.}~\bibnamefont {Wu}}, \bibinfo
  {author} {\bibfnamefont {Y.~X.}\ \bibnamefont {Zhao}}, \ and\ \bibinfo
  {author} {\bibfnamefont {S.~A.}\ \bibnamefont {Yang}},\ }\href {\doibase
  10.1103/PhysRevB.100.041118} {\bibfield  {journal} {\bibinfo  {journal}
  {Phys. Rev. B}\ }\textbf {\bibinfo {volume} {100}},\ \bibinfo {pages}
  {041118(R)} (\bibinfo {year} {2019})}\BibitemShut {NoStop}%
\bibitem [{\citenamefont {Zhang}\ \emph {et~al.}(2020)\citenamefont {Zhang},
  \citenamefont {Wu},\ and\ \citenamefont {Das~Sarma}}]{Zhang2020}%
  \BibitemOpen
  \bibfield  {author} {\bibinfo {author} {\bibfnamefont {R.-X.}\ \bibnamefont
  {Zhang}}, \bibinfo {author} {\bibfnamefont {F.}~\bibnamefont {Wu}}, \ and\
  \bibinfo {author} {\bibfnamefont {S.}~\bibnamefont {Das~Sarma}},\ }\href
  {\doibase 10.1103/PhysRevLett.124.136407} {\bibfield  {journal} {\bibinfo
  {journal} {Phys. Rev. Lett.}\ }\textbf {\bibinfo {volume} {124}},\ \bibinfo
  {pages} {136407} (\bibinfo {year} {2020})}\BibitemShut {NoStop}%
\bibitem [{\citenamefont {Su}\ \emph {et~al.}(1980)\citenamefont {Su},
  \citenamefont {Schrieffer},\ and\ \citenamefont {Heeger}}]{Su1980}%
  \BibitemOpen
  \bibfield  {author} {\bibinfo {author} {\bibfnamefont {W.~P.}\ \bibnamefont
  {Su}}, \bibinfo {author} {\bibfnamefont {J.~R.}\ \bibnamefont {Schrieffer}},
  \ and\ \bibinfo {author} {\bibfnamefont {A.~J.}\ \bibnamefont {Heeger}},\
  }\href {\doibase 10.1103/PhysRevB.22.2099} {\bibfield  {journal} {\bibinfo
  {journal} {Phys. Rev. B}\ }\textbf {\bibinfo {volume} {22}},\ \bibinfo
  {pages} {2099} (\bibinfo {year} {1980})}\BibitemShut {NoStop}%
\bibitem [{\citenamefont {Young}\ and\ \citenamefont {Wieder}(2017)}]{Wieder1}%
  \BibitemOpen
  \bibfield  {author} {\bibinfo {author} {\bibfnamefont {S.~M.}\ \bibnamefont
  {Young}}\ and\ \bibinfo {author} {\bibfnamefont {B.~J.}\ \bibnamefont
  {Wieder}},\ }\href {\doibase 10.1103/PhysRevLett.118.186401} {\bibfield
  {journal} {\bibinfo  {journal} {Phys. Rev. Lett.}\ }\textbf {\bibinfo
  {volume} {118}},\ \bibinfo {pages} {186401} (\bibinfo {year}
  {2017})}\BibitemShut {NoStop}%
\bibitem [{\citenamefont {Wieder}\ \emph {et~al.}(2018)\citenamefont {Wieder},
  \citenamefont {Bradlyn}, \citenamefont {Wang}, \citenamefont {Cano},
  \citenamefont {Kim}, \citenamefont {Kim}, \citenamefont {Rappe},
  \citenamefont {Kane},\ and\ \citenamefont {Bernevig}}]{Wieder2}%
  \BibitemOpen
  \bibfield  {author} {\bibinfo {author} {\bibfnamefont {B.~J.}\ \bibnamefont
  {Wieder}}, \bibinfo {author} {\bibfnamefont {B.}~\bibnamefont {Bradlyn}},
  \bibinfo {author} {\bibfnamefont {Z.}~\bibnamefont {Wang}}, \bibinfo {author}
  {\bibfnamefont {J.}~\bibnamefont {Cano}}, \bibinfo {author} {\bibfnamefont
  {Y.}~\bibnamefont {Kim}}, \bibinfo {author} {\bibfnamefont {H.-S.~D.}\
  \bibnamefont {Kim}}, \bibinfo {author} {\bibfnamefont {A.~M.}\ \bibnamefont
  {Rappe}}, \bibinfo {author} {\bibfnamefont {C.~L.}\ \bibnamefont {Kane}}, \
  and\ \bibinfo {author} {\bibfnamefont {B.~A.}\ \bibnamefont {Bernevig}},\
  }\href {\doibase 10.1126/science.aan2802} {\bibfield  {journal} {\bibinfo
  {journal} {Science}\ }\textbf {\bibinfo {volume} {361}},\ \bibinfo {pages}
  {246} (\bibinfo {year} {2018})}\BibitemShut {NoStop}%
\end{thebibliography}%

\clearpage
\newpage

\onecolumngrid
\appendix

\section{Supplemental Materials for ``$\Z_2$-Projective translational symmetry protected topological phases"}

\section{$\Z_2$ gauge fields in artificial systems}
Let us briefly review how $\Z_2$ gauge fields have been realized in artificial systems, such as photonic/phononic crystals, electric-circuit arrays, and mechanical networks.
\begin{itemize}
	\item In photonic crystals, the sign of coupling between the site rings can be controlled by adjusting the gap between the site ring and the link-ring waveguides, so that a synthetic gauge flux threading each square plaquette can be effectively achieved~\cite{Photonic_Crystal_quadrupole}.
	\item In phononic crystals, positive and negative inter-resonator couplings are achieved by connecting the resonators with thin waveguides on different sides of each resonances nodal line~\cite{Acoustic_Crystal}.
	\item In electric-circuit arrays, negative hopping can be realized by inductors, which can be readily seen from the formula~\cite{Imhof2018,Yu2020}:
	\begin{equation}
	J_{ab}(\omega)=\i \omega (C_{ab}-\frac{1}{\omega^2 L_{ab}})
	\end{equation}
	Here $C_{ab}$ and $L_{ab}$ are capacitances and inductances, respectively.
	\item  In mechanical networks, the negative hopping can be realized by the difference of stiffness coefficients of springs~\cite{Prodan_Spring}.
\end{itemize}

\section{Diagonalization of $\hat{\mathsf{L}}_x$}
In the main text, we diagonalize $\hat{\mathsf{L}}_x$ by a $k_x$-dependent unitary transformation $U(k_x)$, which is explicitly given by
\begin{equation}
U(k_x)=\frac{1}{\sqrt{2}}\left[\begin{matrix}
-e^{\i k_x/4} & e^{-\i k_x/4} & 0 & 0  \\
0 & 0 & e^{\i k_x/4} & e^{-\i k_x/4} \\
e^{\i k_x/4} & e^{-\i k_x/4} & 0 & 0\\
0 & 0 & -e^{\i k_x/4} & e^{-\i k_x/4} 
\end{matrix}\right].
\end{equation}
The Hamiltonian is block diagonalized by the unitary transformation.
It is clear that $U(k_x)$ is not periodic in the Brillouin zone, but $U(k_x)$ and $U(k_x+2\pi)$ are related by
\begin{equation}
U(k_x+2\pi)=U(k_x)V(k_x).
\end{equation}
It is straightforward to derive that $V(k_x)$ is constantly
\begin{equation}
V=-\i \tau_1\otimes\sigma_0.
\end{equation}

\section{Eigenstates}
Let us introduce
\begin{equation}
q(k_y)=J_1+J_2 e^{-\i k_y},\quad m(k_x)=2t\cos\frac{k_x}{2}.
\end{equation}
Then,
\begin{equation}
h_1(\k)=\begin{pmatrix}
m & q^*\\
q & -m
\end{pmatrix}.
\end{equation}
Given $m$, the eigenstate for the valence band is
\begin{equation}\label{Eigenstates}
|\psi_-\rangle  =\frac{1}{\sqrt{2r(r-m)}}\begin{pmatrix}
m-r\\
q
\end{pmatrix},
\end{equation}
with 
\begin{equation}
r=\sqrt{m^2+|q|^2}.
\end{equation}
The wavefunction is clearly periodic for $k_y$ for any $m$. Thus, the Berry connection is derived as
\begin{equation}
\mathcal{A}_2=\langle \psi_-|\i\partial_{2}|\psi_-\rangle=\frac{q\partial_2 q^*-q^*\partial_2 q}{4\i r(r-m)}.
\end{equation}

Therefore,
\begin{equation}
\mathcal{A}_{2}(\k)= \frac{q\partial_2 q^*-q^*\partial_2 q}{4\i r(r-m)}.
\end{equation}



\section{The topological invariant}
Because of Eq.~(24) in the maintext,
\begin{equation}
\sigma_3 h_{1,2}(k_x,k_y)\sigma_3=-h_{1,2}(k_x+2\pi,k_y),
\end{equation}
The conducting and valence eigenstates $|\psi_{\pm}(\k)\rangle$ of $h_1(\k)$ can be chosen to satisfy
\begin{equation}
|\psi_{-}(k_{x}+2\pi,k_y)\rangle=\sigma_3|\psi_{+}(k_x,k_y)\rangle.
\end{equation}
We introduce the Berry phases $\gamma^{\pm}(k_x)$ for each $k_y$-subsystem,
\begin{equation}
\gamma^{\pm}(k_x)=\oint dk_y \mathcal{A}_2^{\pm}(k_x,k_y).
\end{equation}
Here the Berry connection is defined for conducting and valence bands, respectively, as
\begin{equation}
\mathcal{A}_2^{\pm}(k_x,k_y)=\langle\psi_{\pm}(k_x,k_y)|\i\partial_{k_y}|\psi_{\pm}(k_x,k_y) \rangle.
\end{equation}
They satisfy the relation,
\begin{equation}\label{C-V-Connection}
\mathcal{A}_2^{\pm}(k_x+2\pi,k_y)=\mathcal{A}_2^{\mp}(k_x,k_y).
\end{equation}
Recall that for each $1$D $k_y$-subsystem,
\begin{equation}\label{C-V-BerryPhase}
\oint dk_y \mathcal{A}_2^{+}(k_x,k_y)+\oint dk_y \mathcal{A}_2^{-}(k_x,k_y)=2\pi n,
\end{equation}
for some integer $n$. This is because the Brillouin zone $S^1$ of a $1$D insulator can always be regarded as the boundary of a disk $D^2$, and the Hamiltonian can be extended to be an insulating Hamiltonian over the whole disk $D^2$. But as we know, the Berry curvature of the valence bands is opposite to that of the conducting bands. Since the Berry phases are, respectively, the boundary terms of the fluxes, we have the above relations.
From Eqs.~\eqref{C-V-Connection} and \eqref{C-V-BerryPhase}, we derive 
\begin{equation}\label{Berry_Phase_period}
\gamma(k_x+2\pi)+\gamma(k_x)=2\pi n,
\end{equation}
where the superscript `$-$' of the valence-band Berry phase has been suppressed.

For any domain $[k_0,k_0+2\pi]\times [-\pi, \pi]$ in momentum space, we have the identity,
\begin{equation}
\int_{[k_0,k_0+2\pi]} dk_x\oint dk_y \mathcal{F}+\gamma(k_0)-\gamma(k_0+2\pi)=0\mod 2\pi.
\end{equation}
Then, with Eq.~\eqref{Berry_Phase_period}, it gives
\begin{equation}
\int_{[k_0,k_0+2\pi]} dk_x\oint dk_y \mathcal{F}+2\gamma(k_0)=0\mod 2\pi.
\end{equation}
Therefore,
\begin{equation}
\frac{1}{2\pi}\int_{[k_0,k_0+2\pi]} dk_x\oint dk_y \mathcal{F}+\frac{1}{\pi}\gamma(k_0)=0 \mod 1.
\end{equation}
In other words, the topological invariant $\nu$, Eq.~(26) is quantized into integers. Note in the main text we choose $k_0=0$. But in general it can be arbitrarily chosen. Furthermore, because a gauge transformation for the valence band of the $k_y$-subsystem with $k_x=k_0$ can change $\gamma(k_0)$ by $2\pi$, only the parity of $\nu$ is gauge invariant. Hence, $\nu$ is a $\Z_2$ topological invariant.

\begin{figure}
	\includegraphics[scale=1]{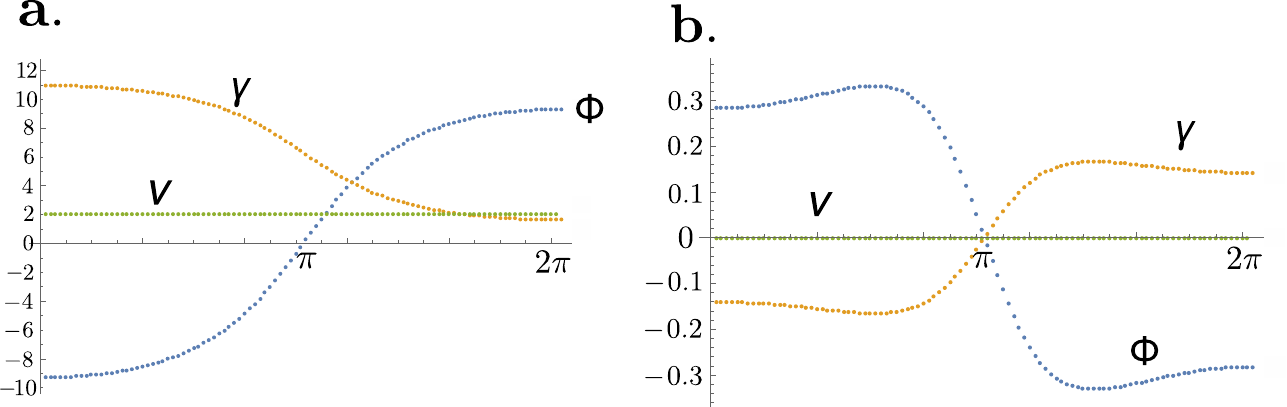}
	\caption{The topological invariant. The $x$-axis stands for $k_0\in[0,2\pi]$. $\gamma$, $\Phi$ and $\nu$ are, respectively, the Berry phase, and the flux of Berry curvature over the domain $[k_0,k_0+2\pi]\times S^1$, and the topological invariant. $\mathbf{a.}$ The nontrivial topological invariant for $t=t_1=1$ and $t_2=2$. $\mathbf{b.}$ The trivial topological invariant for $t=t_2=1$ and $t_2=1$.\label{Topological_Invariant}}
\end{figure}

Particularly, for our system with the valence eigenstate Eq.~\eqref{Eigenstates}, the topological invariant can be straightforwardly calculated as shown in Fig.~\eqref{Topological_Invariant}

\end{document}